\documentclass{JHEP3} 
\usepackage{epsfig}
\epsfclipon

\newcommand{\nco}{\newcommand}
\nco{\beq}{\begin{equation}} \nco{\eeq}{\end{equation}}
\nco{\beqa}{\begin{eqnarray}} \nco{\eeqa}{\end{eqnarray}}
\nco{\lra}{\leftrightarrow}
\def\sfrac#1#2{{\textstyle{#1\over #2}}}

\nco{\sss}{\scriptscriptstyle} \nco{\dphi}{\varphi}
\nco{\lsim}{\mbox{\raisebox{-.6ex}{~$\stackrel{<}{\sim}$~}}}
\nco{\gsim}{\mbox{\raisebox{-.6ex}{~$\stackrel{>}{\sim}$~}}}

\def\wt{\widetilde}
\def\pref#1{(\ref{#1})}

\title{\Large Are Inflationary Predictions Sensitive\\ to Very High Energy Physics?}

\author{C.P.\ Burgess, James M.~Cline, F.~Lemieux\\
Physics Department, McGill University,
3600 University Street, Montr\'eal, Qu\'ebec, Canada H3A 2T8\\
E-mail: \email{cliff@physics.mcgill.ca},
\email{jcline@physics.mcgill.ca}, }
\author{R.~ Holman\\ Physics Department, Carnegie Mellon
University, Pittsburgh PA 15213\\
E-mail: \email{rh4a@andrew.cmu.edu}}

\preprint{McGill-02/33, CMU-HEP-02-11}


\abstract{It has been proposed that the successful inflationary
description of density perturbations on cosmological scales is
sensitive to the details of physics at extremely high
(trans-Planckian) energies. We test this proposal by examining how
inflationary predictions depend on higher-energy scales within a
simple model where the higher-energy physics is well understood.
We find the best of all possible worlds: inflationary predictions
are robust against the vast majority of high-energy effects, but
\emph{can} be sensitive to some effects in certain circumstances,
in a way which does not violate ordinary notions of decoupling.
This implies both that the comparison of inflationary predictions
with CMB data is meaningful, and that it is also worth searching
for small deviations from the standard results in the hopes of
learning about very high energies.}

\begin{document}

\section{Introduction \label{section:intro}}
Decoupling is a double-edged sword. On the one hand it ensures
that most of low-energy physics is independent of the details of
the unknown physics at higher energies, and so allows a meaningful
comparison between experiments and theory before a complete
`theory of everything' is constructed. In retrospect, this was a
prerequisite for the success of physics as a discipline, since it
allows an understanding of Nature at one scale at a time.

On the other hand, decoupling is a real obstacle in our present
situation where we have a candidate theory of everything (string
theory) but only have access to experiments at energies much below
the fundamental string scale. The falsification of such theories
is extremely difficult using any terrestrial experiment we can
currently conceive. This has led many to look for ways in which
very-high-energy physics might have imprinted itself on
cosmological observables which were privy to higher-energy effects
at very early times but with consequences which have survived
until the present epoch.

A recently much-discussed \cite{tp1,tp2,tp3,shenker} candidate for
such an observable is the temperature anisotropy of the cosmic
microwave background (CMB), whose properties are presently being
measured with unprecedented accuracy \cite{boomerang,dasi}. This
particular observable suggests itself as a potential probe of very
high energies because these anisotropies appear to be well
described as arising from primordial metric perturbations (both
scalar and tensor) generated during a much-earlier inflationary
phase of the Universe (see e.g. \cite{liddlelyth}). Given that
such an inflationary phase must have provided at least 60-70
$e$-folds of inflation in order to solve the various
initial-condition problems of the hot big bang, length scales
which are now cosmological in size could easily have been below
the Planck length prior to freezing out as they left the horizon
during inflation \cite{tp1}. This idea was first investigated in a
stringy/quantum gravity context in \cite{tp2}, which also first
pointed out that the effects might be as large as order $H/M$.

Thus, it is not an outrageous hope to think that trans-Planckian
effects could manifest themselves in the CMB, though perhaps in a
subtle way. Several examples of exotic physics which can do so are
put forward in refs.~\cite{tp1,tp2,tp3}, including new particles
having non-Lorentz-invariant dispersion relations, short-distance
modifications to quantum-mechanical commutation relations, and the
preparation of fields in the various nonstandard vacua (more about
which later) that can arise in de Sitter space.

But what of decoupling? Why don't these exotic and hypothetical
high-energy phenomena decouple from the lower-energy physics of
horizon exit? It has recently been argued \cite{shenker} that
decoupling precludes the physics of a high energy $M$ from
contributing effects which are larger than $O(H^2/M^2)$, where $H$
is the Hubble scale when the perturbations of interest leave the
horizon during inflation.\footnote{As pointed out in
\cite{ShiuWasserman}, there is a class of higher-derivative terms
which lead to potentially larger corrections than those described
in \cite{shenker}.} This makes the effects largely undetectable unless $M$ is
very close to $H$. They reach this conclusion by parameterizing
the effects of higher-energy physics at horizon exit using
interactions in a low-energy effective lagrangian, as is known to
be valid elsewhere in other physical applications.

Motivated by the disagreement between the authors of refs.
\cite{tp1,tp2,tp3} and \cite{shenker}, we present here a critical
analysis of how high-energy effects can affect inflationary
predictions. In order to keep our discussion precise we examine a
simple model in which the higher-energy physics is well
understood, consisting of a single heavy scalar particle. By
making the heavy physics mundane and simple, we can explicitly
compute the predictions of the high-energy theory and see when an
effective description applies. In our model a pre-inflationary
period of matter-dominated FRW expansion can occur, which allows us
to unambiguously decide what the initial quantum state of the
system should be. Apart from these reasons, we also believe it to
be useful to know how conventional kinds of high-energy physics
can change the predictions for inflation, not least to use as a
benchmark against which to compare the more exotic types which
have been considered to date.

We find an element of truth in both sides of the above-mentioned
dispute. In a nutshell:

\begin{itemize}
\item We find that it {\em is} possible for the high-energy scalar
to modify the inflationary predictions for the fluctuations in the
CMB, provided the epoch of horizon exit of the relevant wavelengths 
for the CMB does not occur too much
later than the onset of inflation. In fact, by adjusting
parameters, we can have observable effects after as many as 30
$e$-foldings of inflation before horizon exit. The reason this
modification occurs is due to a failure of the adiabatic
approximation stemming from the rapid oscillations of the heavy
field before horizon exit. This then ensures that the epoch before
horizon exit will not be well described by a low-energy effective
theory. This shows how the interference with standard inflationary
predictions is consistent with decoupling, inasmuch as observers
at all times agree on why the high-energy physics does not
decouple.
\item Although we find that the heavy scalar can influence
inflationary predictions, the desired effect is only important in
comparatively special parts of parameter space or for a specific
class of initial conditions. The {\em generic} situation is to
have the high-energy scalar decouple, and so not alter the
standard inflationary predictions. In particular all deviations
from standard inflationary predictions can be made negligible by
having sufficient inflation occur before horizon exit, at least
for the conventional heavy physics considered here.
\end{itemize}

Our results also bear on a related issue which has arisen within
the discussion of trans-Planckian physics. The question is whether
it is {\em in principle} sensible to use the various nonstandard
vacua of de Sitter space. In particular, various recent
calculations have been used to argue that some of these vacua give
physically nonsensical results \cite{dSinconsistent}. We show here
that it is possible to obtain some of the non-standard vacua
starting from the usual vacuum in a pre-inflationary phase of
matter- or radiation-dominated FRW expansion. This possibility
shows that these particular inflationary vacua must make physical
sense, since they arise by time evolving a pre-existing
well-behaved state. The reasonable states which are obtained in
this way differ from those discussed in \cite{dSinconsistent} in
that they approach the standard (Bunch-Davies \cite{bd}) vacuum
for sufficiently high momenta. Together, these results make it
unlikely that reasonable high-energy physics can lead to
nonstandard vacua for modes of arbitrarily high momentum. For
lower momenta, we show that it is always possible to choose a
momentum-dependent nonstandard vacuum in such a way as to
reproduce in the CMB fluctuations the effects we find from a heavy
scalar field.

All told, we find our results to be very encouraging since they
suggest we may be able to have our cake and eat it too. On the one
hand we can sensibly compare standard inflationary results with
CMB measurements, secure in the knowledge that the comparison is
independent of the vast majority of things which could be
happening at higher energies. On the other hand there are
specific kinds of high-energy physics which {\em can} leave their
imprint in the CMB, and for which searches should therefore be
made. In particular we might learn something about how long
inflation had gone on before observed scales left the horizon. The
resulting implications for the CMB might indeed be observable in
the next generation of experiments, such as MAP \cite{map} or
PLANCK \cite{planck}.

We present our discussion in the following way. Section
\pref{S:TheModel} describes our model of high-energy physics. It
consists of a standard hybrid-inflation model \cite{hybrid}
containing a light inflaton, $\phi$, together with a much heavier
scalar, $\chi$. We then examine how the heavy scalar field evolves
if it is not started at a local minimum of its potential, and 
how sufficiently large oscillations of this scalar can
dominate the universe's energy density, causing a period of
pre-inflationary matter domination. The implications of these
oscillations for inflaton fluctuations are described
analytically and numerically in Section \pref{S:Oscillations},
ending with the consequences which they imply for
post-inflationary density fluctuations. Section
\pref{S:Alternative} considers an alternative model, for which
observable implications persist for a longer period during
inflation, and for which the visible signatures might occur at
smaller angular scales than in the previous case. Our conclusions
are briefly summarized in Section \pref{S:Conclusions}.

\section{The Model} \label{S:TheModel}
For the purposes of our analysis we initially adopt a textbook
\cite{liddlelyth} hybrid-inflation model \cite{hybrid}, defined by
the lagrangian density
\beqa \label{eq:modeldef}
    - {\cal L} &=& \sqrt{-g} \Bigl[\sfrac12  \partial_\mu  \phi \,
    \partial^\mu \phi + \sfrac12 \partial_\mu  \chi \, \partial^\mu \chi +
    V(\phi,\chi) \Bigr] , \\
    \hbox{with} \qquad V(\phi,\chi) &=& \sfrac12 \, m^2 \, \phi^2
    + \sfrac14 \, \lambda (\chi^2 - v^2)^2 + \sfrac12 \,
    g \, \chi^2 \phi^2 +  \sfrac{1}{12} \, \tilde\lambda \, \phi^4. \nonumber
\eeqa
The potential has absolute minima at $\chi = \pm v$ and $\phi =
0$, but also has a long trough at $\chi = 0$ provided $g \, \phi^2
> \lambda \, v^2$.  In a later section (3.3) we will consider a variation
of this model with a trilinear coupling, $g' \chi\phi^2$, which
will be shown to have a qualitatively different effect on the CMB
temperature anisotropy.

\subsection{Conditions for Inflation}
In the model (\ref{eq:modeldef}) the inflaton, $\phi$, starts at $\phi = \phi_0$,
with $g \phi_0^2 \gg \lambda\, v^2$, and then slow-rolls along the
$\chi = 0$ trough. To ensure that the trough is sufficiently flat,
we require the following conditions:
\begin{enumerate}
\item 
In order to ensure that $V(\phi,\chi=0) \approx \sfrac14 \lambda
\, v^4$ for all $\phi < \phi_0$ we require $m^2 \,\phi_0^2 \ll
\lambda \, v^4$, as well as $\tilde \lambda \ll m^2/\phi^2_0$
(which allows us to neglect $\tilde\lambda$ in what follows). This
is consistent with having sufficiently large $\phi_0$ provided
$m^2 \ll g \, v^2$. Under these assumptions the Hubble scale
during inflation is $H^2 \approx \lambda \, v^4/(12 \, M_p^2)$,
where $1/M_p^2 = 8 \pi G$. In order to trust Einstein's equations
we take $\lambda \, v^2 \ll M_p^2$, and so have $H^2 \ll v^2$.
\item 
Define the slow-roll parameters \cite{liddlelyth} as $\epsilon =
(M_p \, V'_{\rm tr}/V_{\rm tr})^2$ and $\eta = M_p^2 \, V''_{\rm
tr}/V_{\rm tr}$, where the truncated scalar potential is $V_{\rm
tr}(\phi) = V(\phi,\chi = 0)$. By requiring $4 m^2 M_p^2 \ll
\lambda \, v^4$, or equivalently, $m^2 \ll 3\, H^2$ we can make
the slow-roll parameters small enough  for sufficient inflation.
\end{enumerate}

Under these conditions the field $\chi$ is a heavy degree of
freedom throughout all but the very end of the inflationary epoch.
Its mass is given by the curvature of the potential transverse to
the trough, which depends on the value of the inflaton field
through the relation
\beq \label{eq:Mdef}
    M^2 = \left. {\partial^2 V \over \partial \chi^2}
    \right|_{\chi = 0} = - \lambda\, v^2 + g \, \phi^2 \approx g
    \phi^2.
\eeq
The last approximation follows right up until the end of inflation
by virtue of the conditions described above. We can nevertheless
follow the evolution of this scalar using Einstein's equations
during an inflationary phase provided $\lambda \, v^2 \ll M^2 \ll
M_p^2$. If the couplings $g$ and $\lambda$ are not too different
from one another and are not too small, we also have $m \ll H \ll
M$ during inflation.

\subsection{$\chi$ Oscillations and a Possible Pre-Inflationary Phase}
The picture so far is standard. Our only modification will be to
choose $\chi$ to lie at $\chi_0 \ne 0$, with $\dot{\chi}_0=0$
initially, rather than at the bottom of the trough, $\chi_0 = 0$.
(See \cite{Tetradis} for other discussions of the initial
conditions in hybrid inflation models.) Since large enough $\chi$
oscillations can dominate the energy of the universe, in this
section we do not assume that the Hubble scale $H = \dot a/a$ is a
constant.

If $\chi_0$ is chosen close enough to the trough's bottom ({\it
i.e.} with $\lambda \chi_0^2 \ll M^2$) then we may neglect the
effect of the $\chi^4$ terms in the action on the $\chi$ field
equations, leading to
\beq \label{eq:chieqn}
    \ddot\chi + 3 H \, \dot \chi + M^2(\phi) \, \chi \approx 0.
\eeq
As usual, we consider only the evolution of the homogeneous $\chi$ mode,
i.e. we take $\nabla \chi = 0$, since spatially-inhomogeneous
modes will be redshifted as inflation proceeds.

A general solution to this equation may be obtained, regardless of
the time-dependence of $H$, so long as we may neglect
$\dot\phi/\phi$ and $H$ in comparison with $M$. (We have already
seen this neglect to be justified during the inflationary regime
which is of most interest for the purposes of this paper.)
Assuming the initial condition $\dot\chi_0/\chi_0 \lsim O(H)$ we
find
\beq \label{eq:chiform}
    \chi(t) \approx A(t) \cos\Bigl[M(\phi) (t- t_0) \Bigr],
\eeq
where the slowly-varying envelope is given by $A(t) = \chi_0 \,
[a(t_0)/a(t)]^{3/2}$, and $M^2(\phi)\simeq g \phi^2$. This
evolution describes a fast oscillation rather than a slow roll
because of the condition $M \gg H$, whose origin is discussed
above.

The energy density associated with these oscillations is
\beq \label{eq:chiEnergy}
    \rho_\chi(t) = \sfrac12 \Bigl( \dot\chi^2 + M^2 \, \chi^2
    \Bigr) = \sfrac12 M^2 A^2(t) = \sfrac12 \, M^2 \, \chi_0^2 \,
    \left( {a(t_0) \over a(t)} \right)^3 ,
\eeq
which scales with $a(t)$ as does non-relativistic matter. If the
initial amplitude is chosen to be large enough this energy can be
larger than $\sfrac14 \lambda \, v^4$, in which case the universe
is initially matter-dominated. The amplitude of the $\chi$
oscillations is then damped by the universal expansion, and so
long as the $\phi$ roll remains slow an inflationary phase
eventually begins. As the $\chi$ oscillations damp to zero we
match onto the standard picture of hybrid inflation based on the
slow $\phi$ roll.

In Figs.~(1,2) we plot the total energy density and the $\phi$
kinetic energy for several such rolls, calculated numerically,
where the initial condition $\phi_0$ is fixed but $\chi_0$ varies.
The plot shows how the length of the inflationary period shrinks
as $\chi_0$ gets larger, corresponding to the longer time required
to damp away the energy in $\chi$ oscillations.

\FIGURE{
\epsfig{file=E_density.eps, width=3.5in}
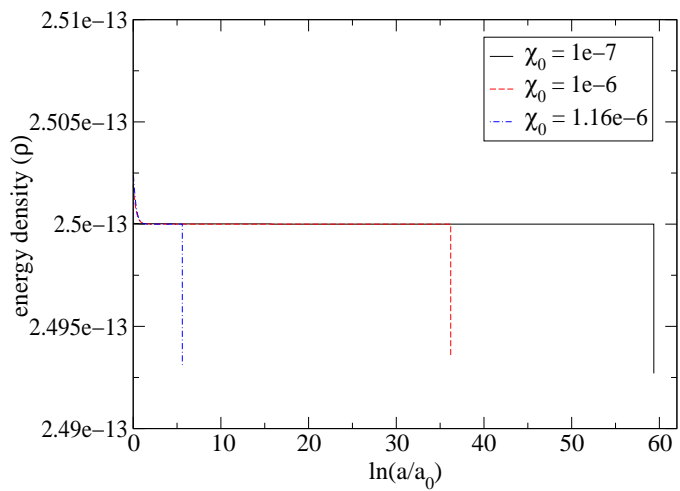
    \caption[Figure 1]{Energy density of the universe
in units of $M_p^4$ for 3
different initial conditions in $\chi_0$, while $\phi_0$ and all
the parameters of the potential are fixed. Here, we have: $\lambda
= g = 1$, $v = 0.001M_p$, $m = 1.1121\times 10^{-7}M_p$ and
$\phi_0 = 0.01945M_p$.}\label{fig:fig1}}

\FIGURE{
\epsfig{file=kineticPart.eps,width=3.5in}
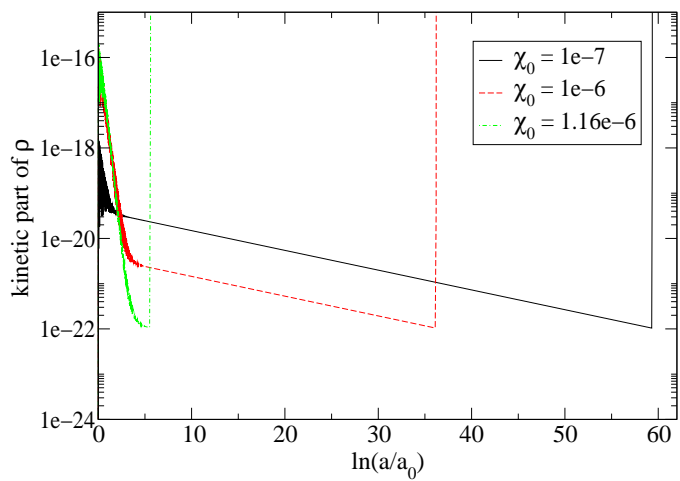
    \caption[Figure 2]{The kinetic energy of the $\phi$ roll for
the same parameters as in the previous figure.}\label{fig:fig2}}

From these considerations we see that there are potentially three
times we should keep track of:
\begin{itemize}
\item We define $t=0$ as the time during inflation when inflaton
fluctuations having $k = k_{\rm\sss COBE} \cong 7.5 H_{\rm
present}$ \cite{lythriotto} leave the horizon: $k_{\rm\sss
COBE}/a_{he} = H$. Here $H_{\rm present}$ denotes the present-day
Hubble scale. We denote the size of the oscillatory envelope at
this time by $\chi_{he} = A_{he} \equiv A(t=0)$. For later
convenience we normalize the scale factor so that $a_{he} \equiv
a(t=0) = 1$.
\item
$t=t_0 < 0$ denotes the time when $\chi$ oscillations begin, which
may or may not lie within a pre-inflationary oscillation-dominated
phase. As in previous discussions $\chi_0 = A_0$ and $a_0$ denote
the values of the $\chi$ field envelope and the scale factor at
this time.
\item
If there is a pre-inflationary phase then $t_i$, with $t_0 < t_i <
0$, denotes the time where inflation begins, as defined by when
the energy in $\chi$ oscillations falls below the constant scalar
potential value $\sfrac14 \lambda \, v^4$. A subscript `$i$' then
denotes the value taken by a quantity at $t=t_i$. If the initial
$\chi$ oscillations are never large enough to dominate the energy
density, we need never consider $t_i$, and we instead define $t_i
= t_0$ and use these two symbols interchangeably.
\end{itemize}

Since the amplitude of oscillations during inflation is
exponentially damped, $\chi_i = \chi_{he} e^{3H|t_i|/2}$, the time
$t_i$ is related to the size, $\chi_{he}$ of the oscillation
envelope at horizon exit. To see this notice that $\chi_i$ is
found by equating the $\chi$-oscillation energy at $t_i$,
$\sfrac12 \, M^2 \, \chi_i^2$ to the inflationary vacuum energy,
$\sfrac14 \, \lambda \, v^4 = 3 H^2 M_p^2$, giving
\beq
    \chi_i^2 = {6 H^2 M_p^2 \over M^2} \, .
\eeq
The number of $e$-foldings between the beginning of inflation and
horizon exit then is
\beq \label{eq:efoldsize}
    H \, |t_i| \sim \sfrac13 \, \ln \left( {\chi_i^2 \over \chi_{he}^2}
    \right) \sim \sfrac13 \, \ln \left( { 6 M_p^2/M^2
    \over \chi_{he}^2/H^2} \right) .
\eeq
Clearly --- for fixed fluctuation size, $\chi_{he}/H$
--- the later horizon exit occurs after the onset of inflation,
the lower $M$ must be, and hence the smaller $m$, $v$ and $H$ must
also be in order to have sufficient inflation {\em after} horizon
exit.

If $t_0 < t_i$ then the initial amplitude must be larger still, by
a matter-dominated expansion factor of $[a(t_0)/a(t_i)]^{3/2} =
t_0/t_i$. The duration of the matter-domination oscillations also
cannot be too long without pushing the initial oscillation energy
above the Planck scale, and so invalidating the use of Einstein's
equations. This condition requires
\beq \label{eq:matterdomlim}
    {t_0 \over t_i} < {M_p \over \sqrt{3} H} \, ,
\eeq
which can be enormous, for small $H$.

\section{Modifications to Inflaton Fluctuations} \label{S:Oscillations}
We now turn to a computation of how the $\chi$ oscillations change
the power spectrum of inflaton fluctuations which get imprinted
onto the CMB. For simplicity, we first consider the case where the
$\chi$ oscillations occur entirely during the inflationary phase
($t_i \sim t_0$).

Consider the quantum fluctuations of the inflaton, $\dphi$. A mode
with wave number $k$ has the equation of motion
\begin{equation}\label{eq:fluc}
    \ddot\varphi_k + 3H\dot\varphi_k + \left[ k^2e^{-2Ht} +V''(\phi)
-g \chi^2(t)\right]\dphi_k = 0.
\end{equation}

In the absence of the coupling to $\chi$, and taking $V''\ll H^2$
as required for slow roll, the standard positive and negative
frequency solutions are
\begin{equation}\label{eq:BDmodes}
     \dphi_k^\pm = (\pm iH + k/a) e^{\pm ik/aH},
\end{equation}
where the scale factor is $a=e^{Ht}$.  In the standard canonical
quantization, it is assumed that the universe starts in the
adiabatic vacuum state which is annihilated by $a_k$ in the mode
expansion
\begin{equation}\label{eq:BDvac}
    \phi \sim \sum_k (a_k \dphi_k^+ + a_k^\dagger \dphi_k^-).
\end{equation}
This is also known as the Bunch-Davies vacuum, $a_k|BD\rangle=0$.
If the universe started in a different state $|\alpha_k\rangle$,
annihilated by some linear combination of $a_k$ and $a_k^\dagger$,
say $b_k = \cosh\alpha_k \, a_k + e^{-i\delta}\sinh\alpha_k \,
a_k^\dagger$ \cite{alpha}, then the mode expansion would be in
terms of functions $\widetilde\dphi_k^+,\ \widetilde\dphi_k^-$
given as linear combinations of the positive and negative
frequencies,
\beqa
    \phi &\sim& \sum_k (b_k\, \widetilde\dphi_k^+ + b_k^\dagger\,
    \widetilde\dphi_k^-),
    \nonumber\\
    \widetilde\dphi_k^+ &=& \cosh\alpha_k\,\dphi_k^+ +
    e^{i\delta}\sinh\alpha_k\,\dphi_k^- .
\eeqa
It is the interference between positive and negative
frequency components of $\widetilde\dphi_k^+$ which gives rise to
potentially observable effects in the CMB power spectrum, since
the latter is derived from $|\widetilde\dphi_k^+|^2$ evaluated at
horizon crossing, $k/a = H.$

If $\delta=0$, the effect on the power spectrum is a modulation of
the standard one,
\beq \label{eq:Pdef}
    \widetilde P_k\sim \lim_{t\to\infty}|\widetilde\varphi_k|^2 =
    |\cosh\alpha_k-\sinh\alpha_k|^2
    P_k,
\eeq
where $P_k$ is the standard Harrison-Zeldovich spectrum.

The simple point we wish to make here is that the coupling of
$\phi$ to $\chi$ can also induce an admixture of positive- and
negative-frequency components in solutions for $\phi$ which are
initially purely positive frequency. In the next section we
demonstrate this explicitly by numerically evolving the fields
forward in time starting from various initial conditions for
$\chi$.

\subsection{Asymptotic Forms}
Before describing these numerical results, we first identify some
of their limiting features by considering an approximate analytic
solution.

In the classical analysis of interest here, the influence of
$\chi$ oscillations on $\phi$ fluctuations can be obtained
explicitly by using the solution $\chi(\phi,t)$ from
eq.\ (\ref{eq:chiform}) in the mode equation (\ref{eq:fluc}). The
mode equations can then be solved perturbatively by use of the
appropriate Greens' function, which since we are interested in the
classical evolution of the fluctuations, is the retarded one. It
is defined by
\beq \label{eq:green}
    G_{k,<}(t,t') = \left\{ \begin{array}{cl} 0, & t<t' \\
            \dphi^+_k(t) \dphi^-_k(t') - \dphi^-_k(t) \dphi^+_k(t'),
    & t>t' \end{array} \right.
\eeq
and obeys the differential equation
\beq
    \left[ \partial_t^2 + 3H\partial_t +k^2e^{-2Ht} \right]G_{k,<}(t,t')  =
    -2ik^3 e^{-3Ht}\delta(t-t').
\eeq

The perturbed positive frequency solution of eq.\ (\ref{eq:fluc})
is
\beqa \label{eq:soln}
    \widetilde\dphi_k^+(t) &\equiv& \dphi_k^+(t) + \delta\dphi_k^+(t),\nonumber\\
    \delta\dphi_k^+(t) &\cong& \int_{t_0}^tdt'\, G_{k,<}(t,t')
    {e^{3Ht'}\over -2ik^3}\, g A^2(t') \, \cos^2[M(t' -t_0)] \, \dphi^+_k(t') ,
\eeqa
and we assume for simplicity no pre-inflationary period, so $t_i =
t_0$. Direct comparison of eqs.~(\ref{eq:soln}) and
(\ref{eq:green}) shows that the perturbed solution has an
admixture of the original positive- and negative-frequency modes.

Since we are interested in the amplitude of the perturbations well
after their oscillations have frozen out, we may take $t\to\infty$
and use eq.~\pref{eq:Pdef} to compute the change in the spectrum
of fluctuations:
\beq \label{eq:dPdef}
     {\delta P\over P}(k) = \left|{\widetilde\dphi_k^+(\infty)\over
     \dphi_k^+(\infty)}\right|^2 - 1 \approx { 2 \; \hbox{Im} \, \delta \varphi_k^+(\infty)
     \over H},
\eeq
where we use the unperturbed result: $\varphi_k^+(\infty) = i H$.

We can use the perturbative expression, eq.~(\ref{eq:soln}), to
deduce useful asymptotic expressions for the solution in the limit
of large and small $k$. Changing variables to $u=(k/H) \,
e^{-Ht}$, the large time limit corresponds to $u\to 0$, and the
integral can then be rewritten as
\beq \label{eq:integralform}
    \delta\dphi_k^+(\infty) =  {H^{2}\over k^{3}}\, g \chi_{he}^2
    \,\int_0^{u_0} {du\over u}
    \, \cos^2\left({M\over H} \ln{u\over u_0}\right) e^{iu} (i+u)
    \left(\sin u - u \cos u\right),
    \eeq
where $u_0 = (k/H) \, e^{-Ht_0}$ and we use $a_{he} = 1$. Taking
first as $k\to 0$ (hence $u_0\to 0$), the integral can be
evaluated to give:
\beq
    \lim_{k\to 0} \delta\dphi_k^+(\infty) \approx -i\,
    {g \chi_{he}^2 \over 18 H} \; e^{3H|t_0|} =  -i\,
    {g \chi_{0}^2 \over 18 H} \, ,
\eeq
up to $O(H^2/M^2)$ corrections. Comparing this to the unperturbed
value $\varphi^+_k(\infty) = iH$, we see that this causes a
suppression of power which depends on the initial fluctuation
amplitude, $\chi_0$, and is independent of $M$ in the limit
$M\to\infty$.

This loss of power at long wavelengths is also seen in the full
(nonperturbative) numerical integration of the equation of motion
(\ref{eq:fluc}), whose results are shown in fig.\ \ref{fig:raw}.
Notice in these numerical examples that the deviation of
$|\tilde\phi_k ^{+} (\infty)|$ from its normal value ($H$) can be
large, even if $\chi_{he}$ is small. A flavor of this also seen in
the perturbative expression due to the conversion of $\chi_{he}$
to $\chi_0$ by the factor $e^{|3Ht_i/2|}$.

\FIGURE{
\epsfig{file=raw-pap.eps, width=3.5in}
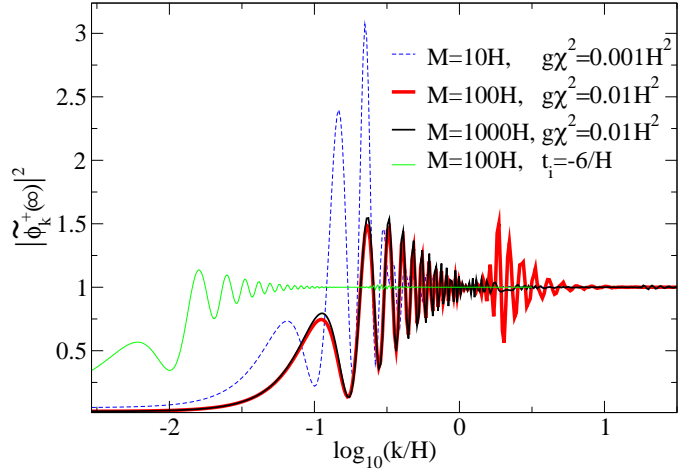
        \caption[Figure 3]{$|\tilde\phi_k ^{+}(\infty)|^2$ in units of $H^2$ for the hybrid inflation
        model, as a function of $\log_{10}(k/H)$
        for several values of $M$ and $g\chi_{\rm he}^2$, rightmost three curves for $t_0=t_i = -4/H$.
Leftmost curve shows the effect of taking an earlier initial time, $t_i=-6/H$, with
$g\chi_{\rm he}^2=10^{-6}$.
}\label{fig:raw}}
\FIGURE{
\epsfig{file=kdiff-pap.eps, width=3.5in}
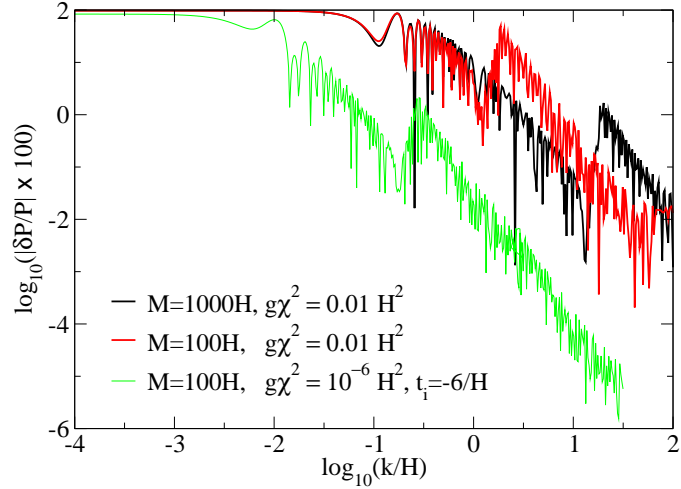
        \caption[Figure 4]{Log of absolute value of percent deviation of power
spectrum as a function of $\log_{10}(k/H)$, for $M=100H$ and
$M=1000H$, with $t_0=t_i = -4/H$ and $g\chi_{\rm he}^2=0.01 H^2$,
and $M=100H$ with $t_i=-6/H$ and $g\chi_{\rm he}^2=10^{-6} H^2$.
Notice that the deviation is large at low $k$ not because the
power is large, but rather because it is smaller than normal.
Order of curves in legends coincides with that at right hand edge
of the graph. }\label{fig:kdiff}}

Eq.~\pref{eq:integralform} also gives the limiting form when
$k\to\infty$ ($u_0\to\infty$), although this limit is somewhat
more difficult to obtain.  Since eq.~\pref{eq:dPdef} shows we are
only concerned with the imaginary part of
$\delta\dphi_k^+(\infty)$ for the power spectrum we focus on this.
In the limit of large $M$ and large $u_0$ we find
\beqa \label{eq:largek}
    \lim_{k\to \infty} {\rm Im }\, \delta\dphi_k^+(\infty) &=&
   {g \chi_{he}^2 \, M \over 4 k^{2}}  e^{H|t_0|} f(k) 
    = {g \chi_{0}^2 \, M \over 4 k^{2}}  e^{-2H|t_0|} f(k), 
\eeqa
where $f(k)$ is an oscillatory function of $k$ of unit amplitude.
Thus the deviations from the standard power spectrum are
oscillatory with an amplitude that falls like $1/k^2$ for large
$k$. This behavior is also confirmed by the numerical results, as
shown in figure \ref{fig:kdiff}.

We can understand the frequency of the fast oscillations in
the numerical results as coming directly from those of the $\chi$
field itself. The numerics indicate that this frequency is the
same as that of the function $\cos[(2M/H)\ln(k)]$.  If we rewrite
$k$ in terms of the physical wavelength of the mode, $k = k_{\rm
phys} e^{-Ht}$, the previous function becomes $\cos(2Mt)$ up to a
phase. This is what we expect for $\delta P$ due to the driving
field, $\chi$, as is seen using the identity $\cos^2(Mt) =
\sfrac12 \, [1 + \cos(2Mt)]$.

The fact that $t_0$ only enters the perturbative results through
the combination $u_0 = (k/H) \, e^{-Ht_0}$ means that the effect
of choosing earlier initial times, $t_0$, to be more negative is
largely to translate the graphs to the left in $\log(k)$ space. At
the same time the normalization changes because of our convention
of using $\chi_{\rm he} = e^{3H|t_i|/2} \chi_0$ as the input
parameter. These features are also borne out by the numerics.

\subsection{Implications for the CMB}
To quantify the effect of the oscillating scalar on the CMB
fluctuations, we have integrated the equation of motion for
inflaton fluctuations numerically for a large range of parameters.
We then compute the fractional deviation in the power spectrum
using eq.~\pref{eq:dPdef}.

In figure 4 we plot the log of the absolute value of the percentage deviation
($\log_{10}(|\delta P|/ P \times 100)$) as a function of
$\log_{10}(k/H)$, for a range of values of $M$, and for two
different values of $t_0$. We find a spectrum of deviations which
oscillates rapidly, due to the fast $\chi$ oscillations, with an
envelope which interpolates between the large- and small-$k$
limits described above.

As can be seen from figures 3 and 4, the largest deviations in the
primordial power spectrum occur at the lowest $k$ values.  This
can be understood because at early times the average value of
$g\chi^2$ is potentially large, and thus effectively gives the
inflaton a significant  mass.  This  in turn gives rise to a
tilted spectrum at low values of $k$, which is constrained by
measurements of the large-angle CMB anisotropy.  We have computed
the effect of input spectra like these on the CMB using CMBFAST
\cite{CMBFAST}, with the results shown in figures 5 and 6.  We
have chosen to normalize all the spectra to a common value (unity)
at $l=10$, and so the depression in fluctuations on very large
scales leads to an enhancement of the size of the first and
subsequent peaks.

\DOUBLEFIGURE[h]{zoomout.eps, width=.51\textwidth} {zoomin.eps,
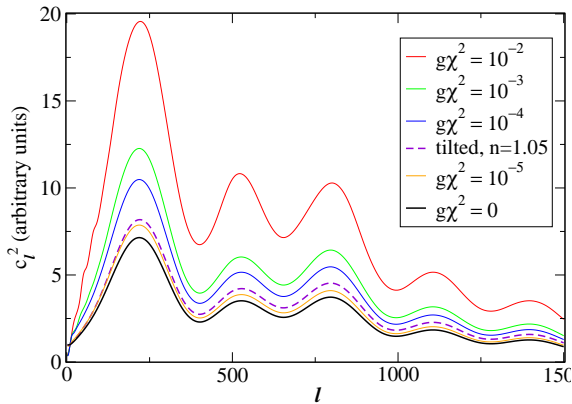
width=.5\textwidth}{Solid lines: Doppler peaks using successively
smaller values of the initial $\chi$ amplitude.  Dashed line:
effect of a tilted spectrum with index $n=1.05$, and no coupling
between $\chi$ and $\phi$. The legend gives $g \chi_{he}^2$ in
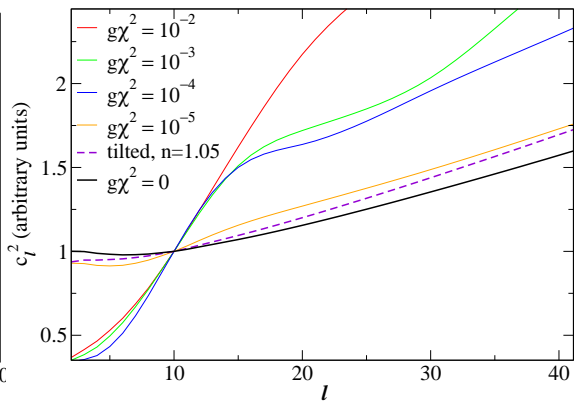
units of $H^2$.}{The same as in fig.\ 5, showing close-up of the
low-$l$ modes. The order of curves in the legend is the same as in
the right hand side of the figure.}

For comparison, the effect of tilting the input spectrum with a
spectral index of 1.05, and no coupling to $\chi(t)$, is shown as
the dashed line in these figures. This shows that the effect of
$\chi$ is qualitatively close to that of tilt, but still in
principle distinguishable.  In particular, the anisotropy of the
tilted model falls below that of the $g\chi_{he}^2=10^{-5}$ model
for $10< l \lsim 70$, but becomes larger for $l\gsim 70$.

\subsection{Phenomenology}
Now that we know how big $\chi$ oscillations must be at horizon
exit in order to be detected we may ask how natural or fine-tuned
our initial conditions must be in order to affect the CMB in an
observable way. Since figures 5 and 6 show that an amplitude $g \,
\chi_{he}^2 \sim 10^{-5} \, H^2$ has implications which are
comparable to a 5\% tilt to the Zeldovich spectrum, we take this
as our benchmark for what an observable deviation would be in the
predictions for CMB fluctuations.

Since the amplitude of oscillations during inflation is
exponentially damped, $\chi_i = \chi_{he} e^{3H|t_i|/2}$,
$\chi_{he}$ will be unobservable if too much time passes between
the onset of inflation and the time of horizon exit. From
eq.~\pref{eq:efoldsize} we see this implies
\beq \label{eq:efoldlim}
    H \, |t_i| \lsim \sfrac13 \, \ln \left( {\chi_i^2 \over \chi_{he}^2}
    \right) \sim \sfrac13 \, \ln \left( { 6 \times 10^{5} \, g M_p^2
    \over M^2} \right) .
\eeq
Choosing all parameters optimistically we could have $g \sim
\lambda \sim O(1)$ and $H \sim 100$ GeV, in which case $v \sim (H
M_p)^{1/2} \sim 10^{10}$ GeV. Taking then $M \sim 10^{12}$ GeV
would give $H|t_i| \sim 14$, so horizon exit can occur quite deep
(more than 10 $e$-foldings) into the inflationary phase.

In summary, we conclude that in this model if inflation persists
for more than about 10 $e$-foldings longer than the minimum
necessary to solve the usual problems of big bang cosmology, none
of the effects which we are considering here are likely to be
observable. However rapid $\chi$ oscillations can significantly
extend the period of inflation before horizon exit to which CMB
observations could be sensitive. The strongest effects on the CMB
anisotropy appear at low $l$ values, and so would not be expected
to be discovered by experiments whose advantage is their smaller
angular resolution. This CMB window onto high-energy physics due
to non-adiabatic oscillations is perhaps no less interesting than
the more nebulous possibility of observing the effects of
trans-Planckian physics.

\section{An Alternative Model} \label{S:Alternative}
For comparison we now turn to a slightly different model, for
which the observable window is wider and has qualitatively
different features. The model is motivated by the observation that
roughly twice as many $e$-foldings would be possible if the
$\chi$-$\phi$ coupling were linear in $\chi$ instead of quadratic:
{\it i.e.} if ${\cal L}_{\chi\phi} = g' \, \chi \, \phi^2$. In
this case the amplitude of $\chi$ oscillations still fall like
$a^{-3/2}$, but this now implies that the perturbations, $\delta
P_k$, to the CMB fluctuations are only damped as $a^{-3/2}$
instead of as $a^{-3}$. Equally interesting is that in this case
the distortions of the primordial power spectrum can have a
maximum at some intermediate value of the wave number, which could
correspond to multipoles of the CMB anisotropy beyond the first
Doppler peak.

We consider, then, a model in which the inflaton field starts near
$\phi=0$ and rolls away from the origin during inflation, such as
the inverted hybrid inflation model \cite{invertedhybrid}.  In
this case $\chi$ is no longer the field which controls the
duration of inflation, whose existence is not important for the
present discussion; rather it is another field which is introduced
just to obtain the new kinds of effects we are interested in
exploring.  The Lagrangian can be written as
\beq \label{eq:newmodeldef}
    V(\phi,\chi) = v^4 - \sfrac12 \, m^2 \, \phi^2
    + \sfrac12 \, M^2 \, \chi^2 + g' \, \chi \,\phi^2 .
\eeq
Here the coupling $g'$ has dimensions of mass. The mass parameter
$M$ in this model does not depend on $\phi$, and we must have
$M^{2}\gg H^2$ to ensure that $\chi$ is not slowly rolling. The
term $v^4$ denotes the constant energy density which dominates
during inflation.

In the limit that the inflaton rolls very slowly so that $\dot\phi$ can be
neglected, the exact solution for $\chi(t)$ is:
\begin{equation}\label{eq:newchisoln}
    \chi(t) = \frac{g'\phi^2}{2 M^2}+\\ \nonumber
   e^{-3 H (t-t_0)/2} \left(\chi_0 -\frac{g'\phi^2}{2
    M^2} \right) \, \cos[{ \wt M}(t-t_0)] \, .
\end{equation}
where $\wt M = \sqrt{M^2 - \sfrac94 H^2}$.  Since we are interested in
the case $M\gg H$, we can take $\wt M\cong M$.

We need to keep the inflaton nearly massless relative to $H$ to insure that it
rolls slowly enough to get sufficient inflation. Since the constant term in
$g'\chi$ is the inflaton mass squared, this requires $g'^2\phi^2/M^2\ll H^2$.
We also demand that  the $\chi$ energy density is subdominant relative to
that of the inflaton; otherwise we initially have a matter dominated era
rather than inflation (a possibility which we will discuss below).  Moreover
we are assuming that $\phi\cong 0$ during inflation, so that the
solution for $\chi$ in eq.\ (\ref{eq:newchisoln}) can be written as:
\begin{equation}\label{eq:chi_sol}
    \chi(t) \simeq  e^{-3Ht/2}\chi_{he} \cos\left( M(t-t_0)\right),
\end{equation}
where $\chi_{he} = \chi_0 \, e^{3 H t_0/2}$.

The calculation of the power spectrum proceeds much as before, and
we only record here those expressions which differ from the
previously-explained hybrid-inflation example. Neglecting the
inflaton mass, the equation governing the quantum fluctuations of
the inflaton, $\dphi$ becomes
\begin{equation}\label{eq:newfluc}
    \ddot\varphi_k + 3H\dot\varphi_k + \left[ k^2e^{-2Ht} 
    - g' \chi(t)\right] \dphi_k = 0 ,
\end{equation}
which generates the same retarded Green's function as before.

The perturbed positive-frequency solution of eq.\ (\ref{eq:fluc})
is then
\beqa \label{eq:newsoln}
    \widetilde\dphi_k^+(t) &\equiv& \dphi_k^+(t) + \delta\dphi_k^+(t),\nonumber\\
    \delta\dphi_k^+(t) &\cong& \int_{t_0}^tdt'\, G_{k,<}(t,t')
    {e^{3Ht'}\over -2ik^3}\, g' A(t')\, \cos[M(t' - t_0)] \, \dphi^+_k(t').
\eeqa
The limiting forms of eq.~(\ref{eq:newsoln}) may be deduced as
before, by changing variables to $u=(k/H)e^{-Ht}$. This time we
find
\beq
    \delta\dphi_k^+(\infty) = - {H^{1/2}\over k^{3/2}}\,g' A_{he}\,\int_0^{u_0}
    {du\over u^{5/2}}
    \, \cos\left({ M\over H} \ln{u\over u_0}\right) e^{iu} (i+u) \left(\sin u - u
    \cos u\right),
\eeq
for which the $k\to 0$ limit then becomes
\beq
    \lim_{k\to 0} \delta\dphi_k^+(\infty) = -i\, {g' A_{he} H\over 2 M^2}\,
    e^{3H|t_0|/2} = -i\, {g' A_{0} H\over 2 M^2}\, .
\eeq
As was true in the previous example, the $k\to 0$ limit of $\delta
\dphi_k^+(\infty)/H$ is a constant, which is proportional to $g'
\, A_0/M^2$.  However in contrast to the previous model, this
deviation vanishes in the limit of large $M$, so that power at
large wavelengths will be relatively unchanged.

The limiting behavior as $k\to\infty$ ($u_0\to\infty$) differs
from our previous example, since the integral converges for large
$u_0$. We find in the large-$k$ limit:
\beq \label{eq:newlargek}
    \lim_{k\to \infty} \delta\dphi_k^+(\infty) \cong 0.5\,{g' A_{he}
    H^{1/2} \over k^{3/2}}\times \hbox{(oscillatory function)}
\eeq
whose amplitude is independent of $M$. These limits are borne out
by our numerical results, presented in figures 7-9, which approach
an $M$-dependent constant for small $k$ and fall with an
$M$-independent envelope for large $k$. The numerical constant in
eq.~\pref{eq:newlargek} is established by comparison with these
numerical results.

\FIGURE{ \epsfig{file=raw2-pap.eps, width=3.5in} \caption[Figure
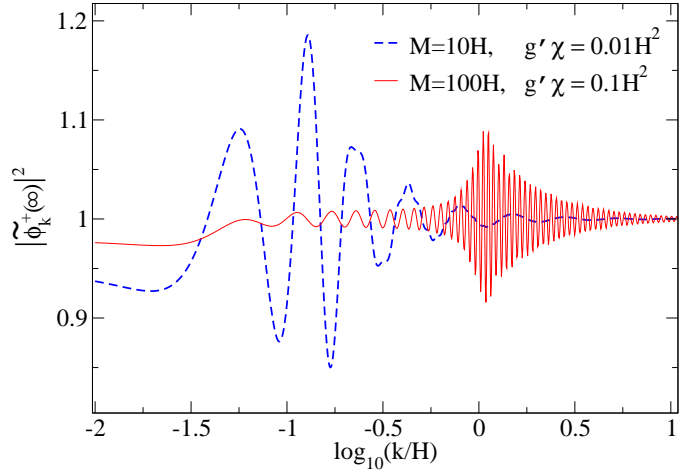
7]{$|\tilde\phi_k(\infty)|^2$ in units of $H^2$ as a function of
$\log_{10}(k/H)$ in the trilinear coupling model, for several
values of $M$ and $g\chi_{\rm he}^2$. }\label{fig:raw2}}

Figures (8,9) plot the log of the percentage deviation
($\log_{10}(\delta P/ P \times 100)$) as a function of
$\log_{10}(k/H)$, for a range of values of $M$, and for two
different values of $t_0$.  The spectrum of deviations is observed
to reach a maximum for wave numbers of order
\beq\label{eq:kmax}
    k_{\rm max} \cong {M\over 2} e^{-H|t_0|}
\eeq
before joining the universal $k^{-3/2}$ fall-off at larger $k$.
{}From these two results we read off the maximum size of the
deviation as being:
\beq\label{eq:Pmax}
    {\delta P\over P}(k_{\rm max}) \cong {g' A_{he} \, e^{3H|t_0|/2}
    \over H^{1/2} (M/2)^{3/2}} = {g' A_{0}
    \over H^{1/2} (M/2)^{3/2}}.
\eeq
This maximum value is therefore greater than the small-$k$ limit
by the factor $(M/H)^{1/2}$.

\DOUBLEFIGURE[t]{fig1a.eps, width=.49\textwidth} {fig1b.eps,
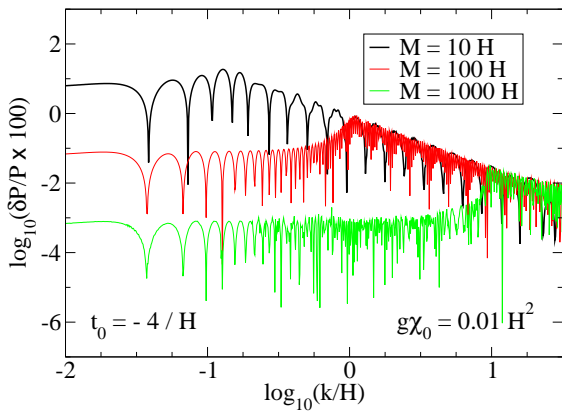
width=.49\textwidth}{The log of the absolute value of the
        percent deviation in the power spectrum
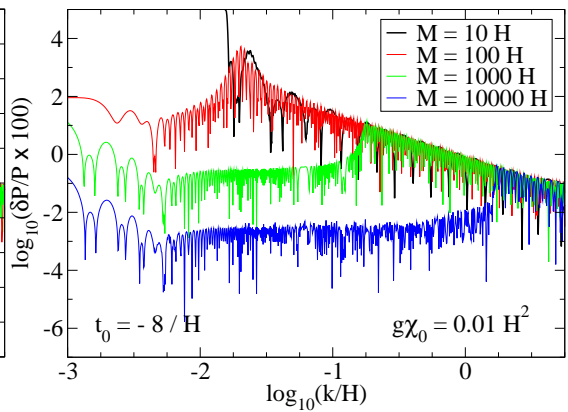
versus log($k/H$) for $t_0 = -4/H$}{The same as in (a) but for
$t_0 = -8/H$. Note that the curve for $M=10H$ goes off-scale for
low $k$ and overlaps with the other curves for high $k$.}

There is a simple intuitive understanding of the above behavior
based on conservation laws and on how amplitudes and wave-numbers
are damped during inflation. The pumping of $\phi$ by the
oscillating $\chi$ field preferentially creates $\phi$ modes with
physical wave numbers which are equal to $M\slash 2$, although the
wave-number is subsequently redshifted. The strongest production
occurs for modes which are pumped right at the beginning of
inflation since the $\chi$ amplitude is largest here, not having
yet been damped as inflation proceeds. This explains both the
value of $k_{\rm max}$ in (\ref{eq:kmax}) and the $k^{3/2}$
scaling of $\delta P/P$ at large $k$. The strongest deviations are
initially created with $k/a(t_0)=M\slash 2$, in agreement with
(\ref{eq:kmax}). Perturbations which are created at later times
$t>t_0$ then correspond to wave-number $k= a(t)M\slash 2 > k_{\rm
max}$, and the effect of $\chi$ on these perturbations is smaller
by the damping factor $[a(t_0)/a(t)]^{3/2} \sim k^{-3/2}$.
Furthermore, we can understand why the deviations also have power
on scales $k < k_{\rm max}$: the pumping amplitude is proportional
to the Fourier transform of the background $\chi$
field. This has an envelope $\sim e^{-3Ht/2}$ in addition to the
oscillations. One therefore expects to see a spectrum of pairs
with $k/a(t_0) \lsim H$.

\subsection{Phenomenology of trilinear coupling model}
With the results of the previous section in hand, we can now
address in more detail what range of parameters could give rise to
a potentially observable effect on the CMB. Since we have already
examined what choices for $H$, $m$ and $v$ maximize the range of
pre-horizon-exit inflation to which the CMB is sensitive, in this
section we instead focus on the $g' \chi \phi^2$ model.

To interpret our results, recall that the scale factor is
normalized such that $a(0)=1$; hence $k$ is the physical wave
number at $t=0$.  The value $k=H$ corresponds to the mode which is
crossing the horizon at that instant. Since we have not specified
how much inflation takes place after $t=0$, we are free to imagine
that this is the value of $k$ which corresponds to some observable
scale of interest, call it $k_{\rm COBE}$.  We are interested in
distortions of the spectrum of perturbations in this region.  The
most fortuitous situation would be that in which $k_{\rm COBE} =
k_{\rm max}$, so that the deviations from the Harrison-Zeldovich
spectrum would drop off for larger or smaller values of $k$,
rather than increase to potentially ruled-out values on one side
or the other.  In what follows, let us focus on this special case,
and consider how much fine tuning of the model parameters would be
needed to get an observable effect.

For the $g'\chi \phi^2$ model, we find the following estimates.
Since we are now assuming that $k_{\rm COBE}=k_{\rm max}=H$, eq.\
(\ref{eq:kmax}) implies that $M = 2H e^{H|t_0|}$, and eq.\
(\ref{eq:Pmax}) then gives ${\delta P\over P}(k_{\rm max}) = g'
A_0 H^{-2}$.  Let us suppose that a $1 \% $ effect is observable;
hence $g' A_0 = 0.01 H^{2}$. Recall the bound that the initial
$\chi$ energy density must be less than that of the inflaton: $M^2
A_0^2 e^{3H|t_0|} \lsim H^2 M_p^2$. Eliminating $M$ and $A_0$, we
obtain a bound on the number of $e$-foldings which can have
occurred prior to horizon-crossing of the COBE-scale mode:
\beq
    H|t_0| \lsim \frac12\ln\left({10^4 g'^2 M_p^2\over 4H^4}\right)^{1/5} \lsim
\frac45 \ln {10 M_p\over H} \sim 33. \eeq
The numerical result 33 is based on taking $g' \sim M_p$ and a
conservative value $H=100$ GeV for the Hubble parameter during
inflation.  Such a value can be obtained for natural choices of
parameters in hybrid inflation \cite{liddlelyth}.  With some
tuning, a lower value of $H$ could be obtained.

This verifies our expectation that the $g' \chi \phi^2$ model is
sensitive to $\chi$ oscillations over more $e$-foldings than is
possible for the $g \chi^2 \phi^2$ hybrid-inflation model. In so
doing it also underlines our observation that trans-Planckian
effects could be mimicked without having to tune the duration of
inflation so that COBE scales crossed the horizon exactly at the
beginning of inflation. An interesting feature of the choice
$g'=M_p$ is that the effect would be caused by relatively small
initial amplitude oscillations of $\chi$, with $A_0
e^{3H|t_0|}\sim 10^4$ GeV. On the other hand, this requires a
small value of the inflaton itself, $\phi < M H/M_p = 0.1$ GeV, to
insure that the inflaton rolls slowly. This again requires the use
of a small-field inflation model like hybrid
inflation\cite{invertedhybrid}.

\FIGURE[t]{ \epsfig{file=mod4.eps, width=5.5in}
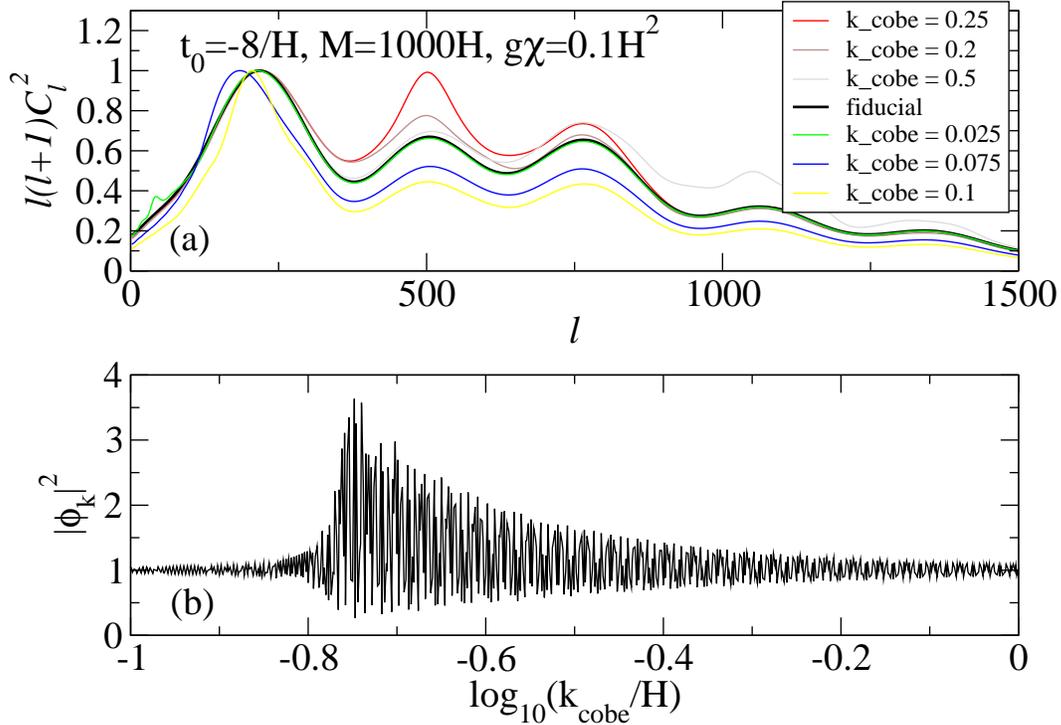
        \caption[Figure 10]{Top: Effect of modulations of the primordial power spectrum on the
        CMB temperature fluctuations.  The order of curves in the legend
coincides with the order in the second Doppler peak.
 Bottom: The modulated input power
spectrum. The scale $k_{\rm COBE}$ is measured in units of
Mpc$^{-1}$.}\label{fig:cmbfast}}

To give a better idea of what the actual observable effects are,
we have used our distorted initial power spectra as inputs to the
CMBFAST \cite{CMBFAST} program, to produce sample plots of the
power spectrum of CMB temperature fluctuations. Figure
\ref{fig:cmbfast} show how a given template for the initial power
spectrum, translates into temperature fluctuations. This template
was produced using the parameters $t_0 = -8/H$, $M=1000H$ and a
large amplitude $g' A_0 = 0.1$, chosen to make the effects more
visible.   The different choices for $k_{\rm COBE}$ (in
Mpc$^{-1}$) correspond to deciding which physical scale to
associate with a given feature in the input spectrum (controlled
by the number of e-foldings between the horizon crossing of that
mode and the end of  inflation).   We have normalized the
different curves so that they all have the same magnitude at the
first Doppler peak.    The fast oscillations of the input power
are washed out in the temperature fluctuations, but the low
frequency modulations have a clearly visible effect.

\FIGURE[t]{ \epsfig{file=mod6.eps, width=5.5in}
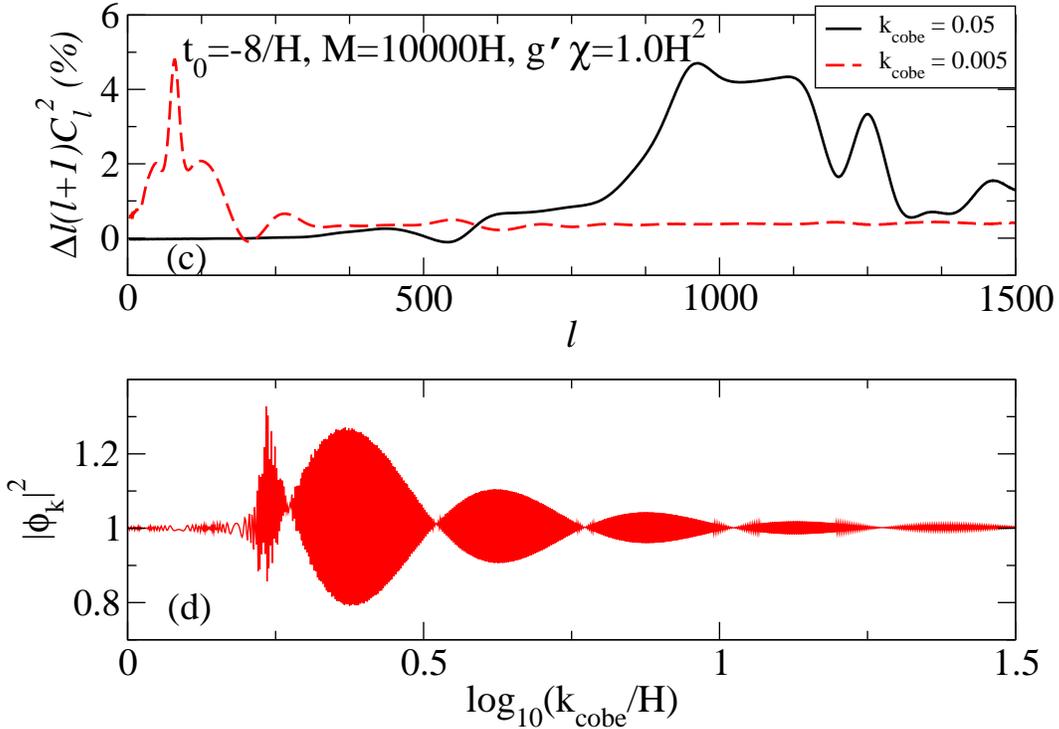
        \caption[Figure 11]{Top: Percent deviation in the CMB temperature
anisotropy due to $\chi$ field oscillations for a given choice of
model parameters. Bottom: The underlying modulated input power
spectrum.  $k_{\rm cobe}$ is in units of Mpc$^{-1}$.}\label{fig:cmbfast2}}

The above examples used an unrealistically large distortion of the
primordial power spectrum in order to illustrate the possibilities
qualitatively.  For a more realistic example, we consider the case
of fig.\ 11 in which the observable effects are at the 5\% level.
To get this large a deviation in the CMB, it is necessary to have
a considerably larger deviation (30\%) in
$|\tilde\phi_k^+(\infty)|^2$, to compensate for the fact that the
oscillatory behavior of the latter tends to wash out its effects.
The effect of shifting the value of $k_{\rm COBE}$ is clearly
visible---it shifts the multipole values where the distortion in
the CMB is most pronounced.  In particular, these examples show
that it is possible to avoid any observable deviation at low $l$.

\subsection{Pre-inflationary Oscillations}

In the previous subsections we took the oscillations of the
pumping field $\chi$ to begin at the same time as the inflationary
phase. We will now show that qualitatively different results can
be obtained by having the oscillations start prior to the
beginning of inflation. In particular the $\chi$ oscillations
could initially dominate the energy density of the universe, so
that inflation is preceded by a matter-dominated period.

The transition from matter to inflaton domination can be modelled
by a scale factor with the behavior
\beqa
    a(t) = \left\{\begin{array}{ll} (t/t_1)^{2/3}, & t < t_1 \\
                e^{2t/3t_1 - 2/3},& t > t_1 \end{array}\right.
    = \left\{\begin{array}{ll} \eta^2/4, & \eta<2 \\
                1/(3-\eta),& 2<\eta<3 \end{array}\right.
\eeqa
Here we have introduced conformal time for convenience, defined by
$d\eta = 2/(3 t_1 a ) d\eta$.  The solution for the oscillating
scalar continues to have the form
\beq
    \chi(t) = \chi_{\rm 0} \cos[ M(t-t_0)] \left({a(t_0)
    \over a(t)}\right)^{3/2} .
\eeq
despite the change in the functional form of $a(t)$ at $t=t_i$.

However in this situation, there is an interesting effect on the
inflaton fluctuations even in the absence of the pumping scalar,
due to the time variation of the background geometry.  The
solutions for the fluctuations have different forms in the matter
and inflaton dominated eras:
\beqa \phi_k(\eta) &=& {A_1} \left(
3\,{\frac {\eta\,k\cos \left( \eta\,k \right) -\sin \left( \eta\,k
\right) }{{\eta}^{4}}}+{\frac {{k}^{2}\sin \left( \eta\,k
 \right) }{{\eta}^{2}}} \right) \nonumber\\ &+&
{A_2}\left( -3\,{\frac {\eta\,k \sin \left( \eta\,k \right) +\cos
\left( \eta\,k \right) }{{\eta}^{4}} }+{\frac {{k}^{2}\cos \left(
\eta\,k \right) }{{\eta}^{2}}} \right), \qquad \eta<2 \eeqa and
\beqa \phi_k(\eta) &=& {B_1}\, \left( k \left( -3+\eta \right)
\cos \left( \eta\,k
 \right) -\sin \left( \eta\,k \right)  \right) \nonumber\\&+&{B_2}\, \left( k
 \left( -3+\eta \right) \sin \left( \eta\,k \right) +\cos \left( \eta
\,k \right)  \right), \qquad \eta>2 \eeqa
It is obvious that if we
form positive and negative frequency combinations of the
independent solutions in each era, there will be mixing of the two
kinds of solutions when we match the functions at $\eta=2$.
Therefore if we start in the vacuum state appropriate for the
matter dominated era, the inflaton will not be in the usual
Bunch-Davies vacuum when it enters the inflationary era.  This
phenomenon was first noticed by Ford and Vilenkin \cite{FV}.
The only scale associated with the spacetime background is the
value of the Hubble rate at the transition time; thus one expects
that modes with $k\lsim H$ will be affected, and there will be no
distortion of the spectrum for $k\gg H$.

This is precisely the kind of behavior we see when we  numerically
integrate to follow the scalar field evolution in the case where
there is no coupling of $\phi$ to an external field $\chi$,
starting with initial conditions appropriate for the purely
positive frequency state of $\phi_k(\eta)$. (We also checked that
the analytic solution given above agrees with our numerical
results.)  The spectral distortion can be clearly seen in the
$g'\chi=0$ curve in Figure 12, where the amplitude of
$\phi_k(\infty)$ is much smaller for $k\ll H$ than for $k\gg H$.
The sign of the effect is surprising; one might have expected
particle production of low-$k$ modes to have produced the opposite
effect.

\DOUBLEFIGURE[t]{etaraw.eps, width=.49\textwidth} {thermal.eps,
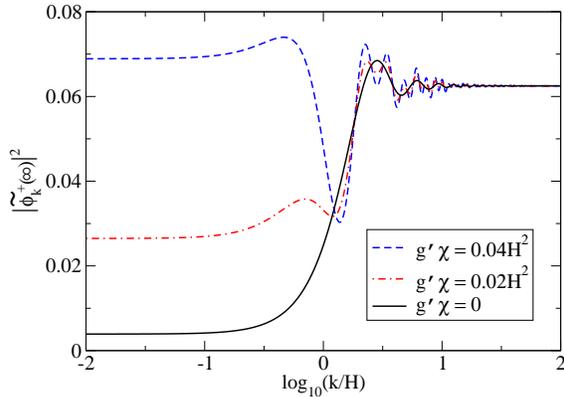
width=.49\textwidth}{$|\tilde\phi_k(\infty)|^2$ as a function of
$\log_{10}(k/H)$ in the trilinear coupling model, when the
universe is initially dominated by the $\chi$ field oscillations.
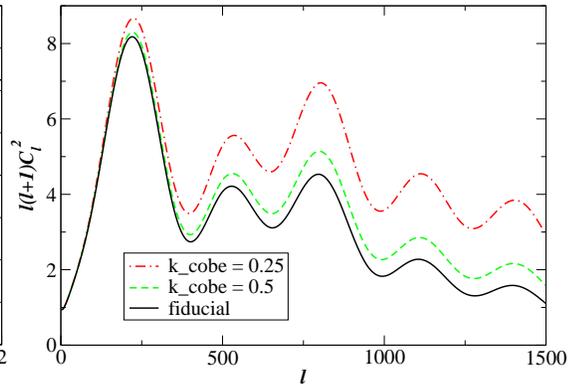
}{CMB temperature anisotropies for the $g'=0$ case, compared to
the standard best fit cosmological model.}
As before, by adjusting the total duration of inflation we can
shift the effect of the distortion to lower or higher multipole
values in the CMB, as illustrated in figure 13.

When we turn on the coupling $g'$, we see from figure 12 that it
is possible to counteract the suppression of power at low $k$.
However it would require some fine tuning of $g'$ to arrange for
this compensation to be exact.  In any case, the new features in
the spectrum occur for $k$ values in the range $k\lsim H$.  The
only way these can show up in the CMB is if the transition between
matter and radiation domination occurs just when the relevant
modes cross the horizon.  Thus we are back to the situation where
the amount of inflation is the bare minimum compatible with
solving the horizon and flatness problems.

This is in contrast to the previous case where horizon crossing
could occur some time after the beginning of inflation by
adjusting the $\chi$ mass $M$ to larger values.  The difference is
that, similarly to the standard hybrid inflation model, in the
present case the largest distortions of power always occur at
small wave numbers.  Although there is a peak in the spectrum at
high $k$ given by $M e^{-H|t_i|}/2$, it is only a local maximum,
and never exceeds the size of the deviation that occurs at low $k$
values.  This is the essential difference between the present
situation and the previous case where the evolution was presumed
to start during inflation.  To reiterate, the difference arises
because the prior period of matter domination puts the inflaton in
a state which is not the same as the usually presumed Bunch-Davies
vacuum.

\FIGURE{\epsfig{file=eta_new2.eps, width=3.5in}
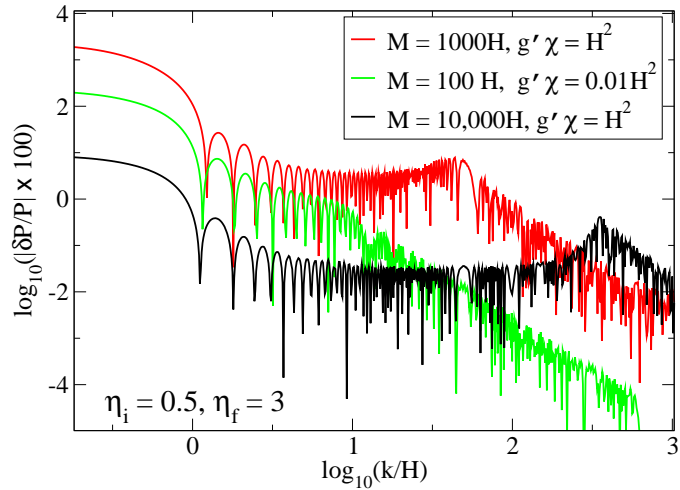
        \caption[Figure 3]{Log of
the percent deviation in the power spectrum versus $\log(k/H)$
when inflation is preceded by a matter dominated period.  Order of
legends is the same as that of curves at the left hand side of the
graph.}\label{fig:matter}}

\section{Conclusions} \label{S:Conclusions}
Clearly, using the CMB to probe energy scales so far beyond what
could be done in accelerators in an intriguing possibility. But,
at least in principle, this approach carries the risk of
completely undermining the predictiveness of inflationary models,
and so negating the value of comparing them with CMB observations.

What our results show is that there can be a reasonable window
onto higher-energy physics purely within a model under which all
calculations are completely under quantitative control. The window
we find is onto any epoch of inflation prior to horizon exit but
after inflation begins, provided that this epoch is not too long
({\it i.e.} not longer than from 10 to 30 $e$-foldings, depending
on the model). A longer pre-horizon epoch of inflation would be
sufficient to completely remove the effects for the CMB of any
early oscillating scalar modes.

Indeed, it is not too surprising that some sensitivity exists
within the CMB towards the cosmic expansion just before horizon
exit, since the standard inflationary predictions relate the CMB
fluctuations to the size of the slow-roll parameters at horizon
exit. For instance, if observed scales were to exit the horizon
just as inflation began, then even standard calculations would not
predict a spectral index near unity. What our calculation does is
to extend this sensitivity to many more $e$-foldings of
pre-horizon-exit inflation provided that the non-adiabatic scalar
oscillations we consider are present during these times.

Three features of our results bear particular emphasis.
\begin{itemize}
\item
First, they allow for new effects in the CMB, but do not make them
generic. This means that the standard predictions are very robust
to higher-energy physics, just as they had been expected to be.
But our results also encourage the search for deviations from
standard predictions, since they broaden the kinds of physics
which can plausibly be expected to produce them.
\item
Second, our results do not contradict any of the well-established
implications of decoupling. The kinds of oscillations we consider
would not be expected to decouple by observers at any energy under
consideration, since they can continually pump energy $E \sim M$
into the lower-mass modes. All observers therefore understand why
it would be wrong to try to integrate out the heavy fields having
mass $M$ in the usual way, since modes of energy $M$ are
constantly being pumped by the scalar oscillations.
\item
Third, it is clear from our calculations -- as well as from older
calculations \cite{FV} -- that for low wave-numbers the $\alpha$
vacua of de-Sitter space do make physical sense, since they may be
obtained by evolving the usual FRW vacua forward in time in a
non-adiabatic way.  $\alpha$-vacua can be expected to be obtained
in this way up to a maximum wave-number, $k_{\rm max}$, whose
value is set by the most energetic states which can be excited by
the non-adiabatic evolution. Recent objections in principle
\cite{dSinconsistent} to using $\alpha$ vacua may in this way be
seen to be traceable to the use of these vacua up to arbitrarily
high wave-numbers.
\end{itemize}

Our calculation does not directly preclude the possibility that
trans-Planckian physics might yet contribute uncontrollably to
inflationary predictions to the CMB, since we have restricted our
focus entirely to sub-Planckian physics for which calculations can
be systematically made. Indeed it is likely to remain difficult to
draw any definitive conclusions on this score without having a
detailed understanding of the nature of trans-Planckian physics.

However, we believe our calculation {\em does} a new
features to the trans-Planckian debate. First, it identifies
non-adiabaticity as a key ingredient of previous examples of how
trans-Planckian effects can influence the CMB. (This is in
agreement with the results of Martin and Brandenberger \cite{tp1},
but by disentangling the adiabaticity from more exotic possibilities
like non-Lorentz-invariant dispersion relations we also show that
these exotic assumptions are not crucial for obtaining an
observable effect.) The relevance of adiabaticity can be traced to
its role in preventing the heavy physics from being integrated out
despite the modes involved being heavy. From this point of view we
believe that if trans-Planckian physics obeys our usual notions of
decoupling (as, for example, string theory appears to do) then it
is unlikely that it can affect the CMB unless it introduces new
low-energy states or non-adiabatic physics at the scale of horizon
exit.

Second, in the event that controlled calculations of
trans-Planckian physics should ultimately be possible, our
calculation provides a useful benchmark against which they may be
compared. Indeed such a comparison would be necessary in order to
distinguish really new trans-Planckian physics from the more
mundane types which we consider here. In particular, our
calculation shows that it is always possible to mock up an
observed modulation of the observed CMB power spectrum in terms of
some sort of $\alpha$-vacuum, so in itself the power spectrum
cannot be used to tell if any observed deviations are due to an
$\alpha$-vacuum or some other physics.

At any rate, it is probably safe to say that inflation offers the
best hope to find relics of very-high-energy physics at the
present time, be it trans-Planckian or not.

\acknowledgments

The authors would like to acknowledge the Aspen Center for Physics
where this work was begun. We thank G.\ Shiu for helpful comments on
the manuscript, and U.\ Seljak for his kind assistance with
CMBFAST.  We also thank R. Brandenberger, B. Greene and S.~Shenker
for stimulating discussions.  R.~H. was supported in part by DOE
grant DE-FG03-91-ER40682, while the research of C.B., J.C. and F.L.
is partially supported by grants from N.S.E.R.C. (Canada) and
F.C.A.R. (Qu\'ebec).

\end{document}